# UAV Based 5G Network: A Practical Survey Study

Mohammed Abuzamak, Hisham Kholidy
State University of New York (SUNY) Polytechnic Institute, Network and Computer Security Department, Utica, NY USA
abuzamm@sunypoly.edu, hisham.kholidy@sunypoly.edu

*Abstract*- **Unmanned aerial vehicles (UAVs) are anticipated to significantly contribute to the development of new wireless networks that could handle high-speed transmissions and enable wireless broadcasts. When compared to communications that rely on permanent infrastructure, UAVs offer a number of advantages, including flexible deployment, dependable line-of-sight (LoS) connection links, and more design degrees of freedom because of controlled mobility. Unmanned aerial vehicles (UAVs) combined with 5G networks and Internet of Things (IoT) components have the potential to completely transform a variety of industries. UAVs may transfer massive volumes of data in real-time by utilizing the low-latency and high-speed abilities of 5G networks, opening up a variety of applications like remote sensing, precision farming, and disaster response. The collection of processing of data from multiple sensors is made possible by the incorporation of IoT devices in UAVs, enabling more precise and effective decision-making. In conclusion, by enabling new degrees of automation and connectivity, the integration of UAVs and 5G technology with IoT devices have the ability to fundamentally alter how we work and live. This study of UAV communication with regard to 5G/B5G WLANs is presented in this research. The three UAV-assisted MEC network scenarios also include the specifics for the allocation of resources and optimization. We also concentrate on the case where a UAV does task computation in addition to serving as a MEC server to examine wind farm turbines. This paper covers the key implementation difficulties of UAV-assisted MEC, such as optimum UAV deployment, wind models, and coupled trajectory-computation performance optimization, in order to promote widespread implementations of UAV-assisted MEC in practice. The primary problem for 5G and beyond 5G (B5G) is delivering broadband access to various device kinds. Prior to discussing associated research issues faced by the developing integrated network design, we first provide a brief overview of the background information as well as the networks that integrate space, aviation, and land.**

## I. INTRODUCTION

The Fifth Generation of Mobile Communications (5G), the next step within the development of telecommunication networks, is marked by a profound transformation of both the infrastructure as well as the technologies used. The number of users/devices demanding Internet access with various performance standards, as well as the variety of apps and use cases, will increase with the adoption of 5G. One of these more modern use scenarios is the Internet of Things (IoT), which will include a greater amount of interconnected devices and be used in a variety of application situations [2]. Future fifth-generation (5G) radio access networks are anticipated to effortlessly and widely connect everything in their environment. The Internet of Things (IoT) is currently very popular, which has led to an increase in mobile data traffic for the next 5G and beyond 5G (B5G) wireless networks. It would result in significant capacity demands on the existing infrastructure and place a significant financial and operational load on telecom providers. Due to high operational costs and complex, unstable conditions, the deployment of terrestrial infrastructures is difficult and economically unviable. This is where unmanned aerial vehicles (UAVs) come to help create a solution to this everlasting problem. Drones have been looked at as a promising paradigm shift to enable three main use scenarios for future wireless devices, namely improved mobile broadband (eMBB) plus internet communications (mMTC) and hyper-low-latency communication. (URLLC) [1]. This is one of the main reasons for UASs being an extremely important aspect of 5G. Due to their inherent mobility, UAVs need wireless capability for connectivity. Additionally, the overall quality of experience (QoE) inside the area can be improved by combining UAVs with the forthcoming 5G mobile system for users and apps that depend upon low latency and increased throughput. For telecommunications networks, a mix of these two techniques may be advantageous. For instance, the low power requirements of 5G activities may assist with the UAV side power constraint, while networking can be made more secure by using encryption in 5G. UAV scans to aid in planning in areas with only older generation networks, which have congestion difficulties, and in the implementation of 5G in areas wherein the company is engaged will be sluggish or nonexistent. Licensed or unlicensed bandwidth can be used to supply the wireless communication infrastructure. The licensed spectrum functions for UAVs can be carried out in a number of methods, including using current cellular bands, satellite technologies, or distinctly licensed spectrums set aside for UAVs. Unlicensed bandwidth is owned by numerous parties and is subject to additional situations of disturbance and competition. In contrast, licensed spectrum necessitates decisions and offers dependable channel allocation for UAV communications. Low-altitude UAVs are widely employed for a variety of uses and purposes in a wide range of fields due to their adaptability and great mobility. UAVs can be used as airborne networking sites from the perspective of wireless communication issues. UAVs are assuming significant societal responsibilities, particularly in the area of communications, in which they are the go-to option when temporary assistance and a favorable cost-benefit ratio are needed. Work efforts like this aim to reduce the price of UAV batteries, which boosts effectiveness and uptime. To maximize energy efficiency, one of the most popular solutions to this issue is to discover the UAV's ideal placement and trajectory. To cut down on the amount of energy used by UAV BSs, a technique for monitoring a power path is suggested. By satisfying the dependability and delay needs of IoT applications and consumers in difficult-to-reach places, this solution enhances UAV communication and mobility. To be able to significantly minimize the rapid rise in demand which is one of the main issues causing congestion during hotspot events, finish the established mobile data framework, offer additional wireless connectivity assistance to obstructed users, expand the portable system's capacity, reduce coverage gaps, and finish the established mobile network structure. UAVs can offer moment-on-demand services in contrast to established infrastructures because of their flexibility, quick deployment without restrictions, and low cost in comparison to traditional services. UAVs can be used for net backup to increase fault tolerance and to provide communications to places with no infrastructure. UAVs can be employed as the main piece of machinery to support rescue efforts as well as a communication tool to speed up self-rescue amid areas prone to environmental disaster. Overall, by allowing new levels of automated efficiency, and connectivity, the combination use of UAVs alongside 5G networks with IoT devices does have the potential to completely alter a variety of industries. In the years to come, it's possible that we will witness even more fascinating breakthroughs in this area as technology develops.

## II. LITERATURE REVIEW

One of the main ways we can implement UAVs into our society in a beneficial way is through the use of disaster control with UAV devices and 5G. IoT-enabled UAVs can be directed to fly over a damaged region while gathering relevant data, including films, images, gas levels, and much more. The emergency workers will gain from the data gathered because it will assist them to arrive in this location well-prepared [2]. Also, it could be beneficial because it can transfer things such as water, medications, and even food. According to the guardian flying drones were sent to help deliver things like food and water and medication and test whether conditions are beneficial or not (Guardian). To give a couple more examples of UAVs helping to rescue when there's a disaster when the COVID-19 pandemic happened in 2020 certain areas were hard to access because nature purposes and helping those people would take a lot longer than the time needed to get to them. In May 2020, In the isolated Scottish Highlands, COVID-19 test kits and PPE supplies were delivered to persons in need using drones. Drone deliveries substantially cut delivery times from up to six hours to only 15 minutes (Adorama). Another thing that UAVs can be used for is surveillance. It would be an amazing advantage to have surveillance over your city because there would be fewer cops who need to be out there risking their lives doing daily surveillance when we can have an IoT device doing 24/7 surveillance and all of that data goes back to the police station. Another way UAVs are being developed today is to create robots that are armed with guns and that can go out on the battlefield instead of humans having to risk their lives in war. These are only some examples of how UAVs can help our society today. The SUAP mechanism helps mobile networks function better during emergencies. To acquire the required online services as well as provide interim connectivity to boost system performance, the positioning of UAV-BS was represented as an optimization issue. A route-planning technique was created in the work for UAVs placed in regions with spotty network connectivity for data-streaming applications. In contrast to conventional planning, this calls for the UAVs' QoS to maintain bandwidth. In order to develop a combinatorial optimization issue that aims to reduce the flying costs of a UAVs over the area of interest, a mathematical model was given. By estimating the amount of UAVs and mapping out their flight paths, the heuristic method reduces energy consumption and boosts link throughput. Similar to this, the authors suggest deploying a group of UAVs known as AAPs in three dimensions for downlink communication while taking inter-cell noise and energy usage into account. This same placement issue has been split into two components the lateral one, which would be critically derived from of the energy usage of the AAP and the inter-cell interference, as well as the second section, this same lateral positioning, which is handled with a spherical packing problem and solved by a multi layered frequent polyhedral method which provides the maximum this same compressibility of a service area of APP and also defines a required arbitrary cap on the number of APPs that is utilized [3]. Problems that are really challenging to solve can be solved in real time. In develops an effective technique for UAV location in terrain with actual impediments. Depending on the typical real-time situation, the technique can be adjusted for line of sight. The research establishes that accurate global positioning is adequate to identify the target area's linear search path. Comparing the approach to tactics based on probabilistic models, it demonstrated higher performance benefits. The authors suggested an automated approach for 3D UAV positions for real-time issues without the usage of GPS or other observable mobile signals supplied by the UAV [3]. The technique makes advantage of current mobile infrastructure to let the UAV pinpoint the locational on the placement of four nearby base stations. The best border was found using heuristics as a solution. In contrast to existing strategies like optimization, which have trouble operating in real time, this methodology employs learning algorithm and optimization algorithms to address the issue in instantaneously, which is crucial for such applications. For the purpose of achieving UAV ego in a wireless transmission system, a type of cognitive learning is proposed in. In order to obtain flexibility in the immediate motions of UAVs and make appreciable gains in key performance measures, the approach leverages the concept of parameter mutation mixed with virtual pressure. The work of suggests a method based on UAV-BSs to reduce the time it takes to identify power failures inside the power network and to provide on-demand coverage. The idea deploys wireless communication devices taking into account Wi-Fi backhaul and user demand after using UAV-BSs as well as a Q-learning system to automatically identify gaping holes. In in order to increase the number and connected users and improve the 3D positioning of UAV-BS at various angle points, machine learning helped the UAV BSs can remove the human-in-the-loop (HiTL) model.

### *A. UAV/5G PHYSICAL LAYER*

UAV-assisted communication networks are now the subject of many works, particularly in unanticipated or transient occurrences. UAV communications' effectiveness can lead to universal support and service huge flexible interconnections thanks to the portable transceiver capability and cutting-edge signal processing methods. The case when UAVs serve as hovering BSs (UAV-BSs) is shown in Figure 1. These UAVs have been typically outfitted with a variety of malware for accepting, handling, and conveying signals, also with the goal is to complete cellular wireless systems by adding the extra ability to spot all through power fluctuation.

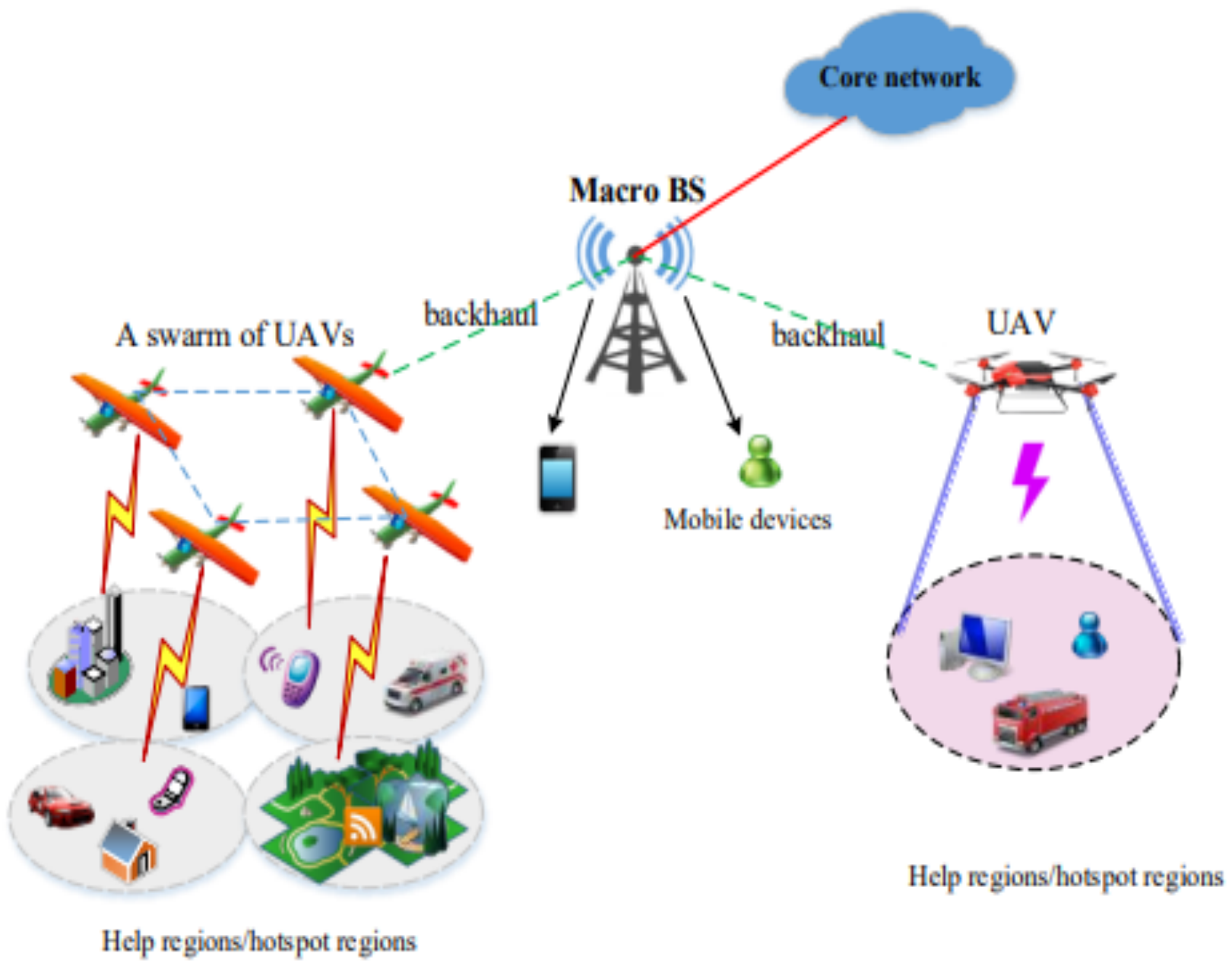

***Fig. 1: UAVs Serve As Hovering BSs***

Its usage of UAVs for aerial BSs can serve as a designated target is illustrated in this example. Every UAV is outfitted with wireless transceivers that enable communication with both base stations as well as other UAVs. Physical layer approaches are a major concern since they have a significant impact on UAV applications, in order to improve the overall performance of the system for UAV communication on 5G networks. Transmission over mmWave, NOMA transfer, CR, and fuel cells occur at the physical layer are the core five possible critical technologies [1]. This is important to keep in mind that UAVs may need to handle many data kinds, including voice, video, and large data files that present unheard-of issues in terms of high bandwidth requirements.

### *B. UAV/5G NETWORK LAYER*

In order to create an inter architecture, such as unmanned aerial systems levels for huge broadcast service areas, ground cellular levels of D2D communications for mobile users, tiers for limited radio coverage locations, and so forth the next-generation networks must smartly and painlessly combine various nodes. The examination of network layer approaches will face new challenges as a result of the integration of several tiers. As a result, particular QoS-coordination techniques are required. These operators are put under an unreasonable amount of financial and operational pressure by the unceasing growth in mobile traffic volumes. The installation of terrestrial infrastructures was difficult in unexpected or brief situations, nevertheless, because of how complex and dynamic movable environments were. A solution that would be helpful for new areas which are in demand is to improve the QoS of the ground users by bringing them closer to the drone cells because of the short-range LoS communications as shown in Figure 2.

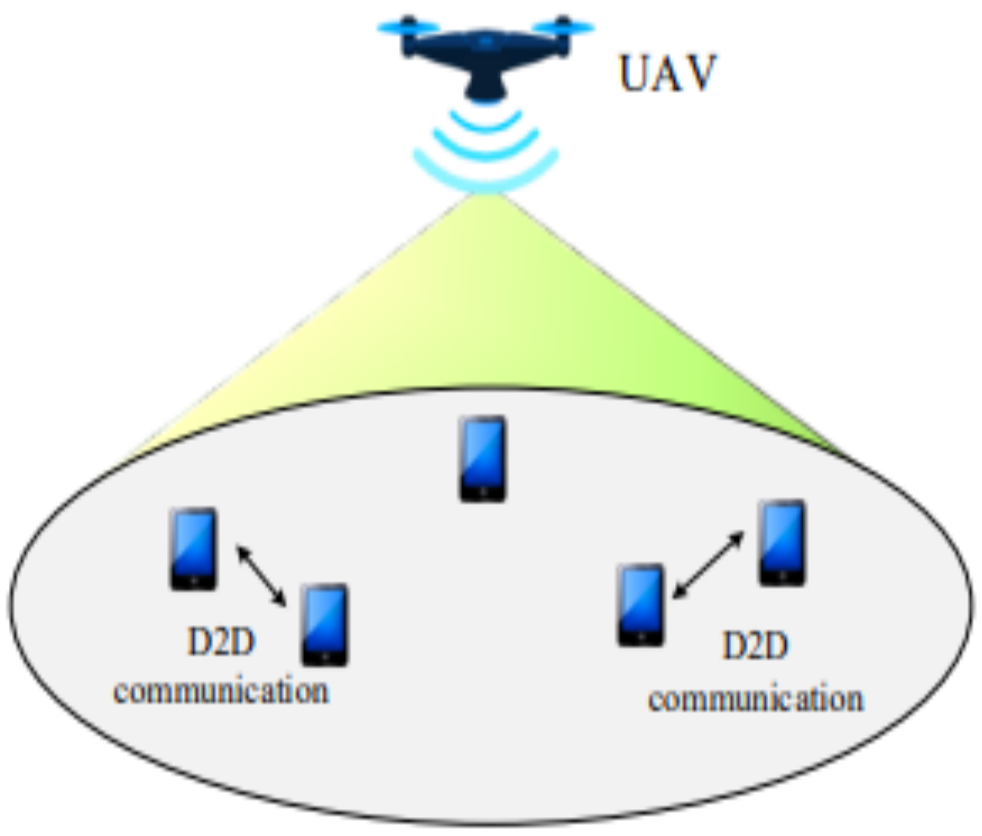

***Fig. 2: Long-Range IoT***

Each layer of the model with several UAVs was analyzed separately. For the precise and effective positioning of the UAVs, primary concern domination and volatility techniques were combined in a clever way suggested [1]. Focused on examining the viability of multi-tier UAV network architecture beyond limited lifetime UAV systems recognized the relevant challenges, including such UAV energy consumption, in perspective of the bandwidth utilization of a downlink transmission and disturbance control. Finally, numerical outcomes demonstrated the influence of various urban contexts on our multi-tier UAV network architecture.

### *C. UAVs IoT & 5G*

5G, the fifth generation of mobile network technology, is one of the innovations that have the potential to change the use of drones. The usage of 5G in UAVs has the potential to significantly improve their capabilities and open up a plethora of new applications. For example, the faster speeds and reduced latency of 5G networks enable full video transmission from unmanned aerial vehicles (UAVs), which may be utilized for applications including such searches and rescue, surveillance, and infrastructure inspection. Because of the improved capacity of 5G networks, significant volumes of data, such as high-resolution photos and video, may be transmitted and utilized for mapping and data collecting. One of the primary advantages of adopting 5G in UAVs is the capability to operate beyond the operator's line of sight (BLOS). Traditional unmanned aerial vehicles (UAVs) rely on a radio system between both the UAV as well as the operator, limiting its reach to the range that the controller can see. UAVs may be linked to the internet and operated remotely via 5G, allowing them to fly outside the operator's line of sight and accomplish jobs in inaccessible regions. Another possible 5G use in UAVs is the usage of swarms, in which several UAVs collaborate to execute tasks. UAVs inside a swarm could coordinate and communicate their operations in real-time thanks to the fast speeds and low latency of 5G networks, allowing them to perform tasks like cartography, surveillance, and rescue missions more effectively. The restricted reach of a 5G network is one of the problems of implementing 5G in UAVs. While stations or transmitters can be used to enhance the reach, these options are not always feasible or cost-effective. Another difficulty is interference inside the high-frequency band used by 5G, which might hinder the performance of UAVs. Despite these obstacles, the deployment of 5G in UAVs does have the potential to dramatically expand their performance and allow a broad range of new applications. As technology advances, we may expect to see more drones outfitted with 5G technology, as well as increased usage of 5G in UAVs. Here are some more of the potential uses of 5G in UAVs.

- Infrastructure inspection and maintenance, boosting safety and minimizing the number of physical inspection teams
- Medical equipment and supplies are delivered to rural or underdeveloped locations, enhancing access to healthcare.
- Actual video transmission and operation outside line of sight.
- Disaster response, animal monitoring, and weather forecasting, as well as data collection and transmission from a variety of sites
- Swarms are used for activities including mapping, surveillance, and search and rescue missions.

Having previously said, there are various approaches to integrating UAVs/satellites with the 5G domestic network in order to serve the IoT use case and doing so brings about a variety of advantages. If there is no other communication infrastructure, UAVs and satellites may be used in 5G application areas with the goal of gathering and transmitting the data produced by IoT devices. By serving as mobile communication sites where position and motion routes can be appropriately configured and dynamically altered, UAVs can aid in extending the scope of corporate IoT networks

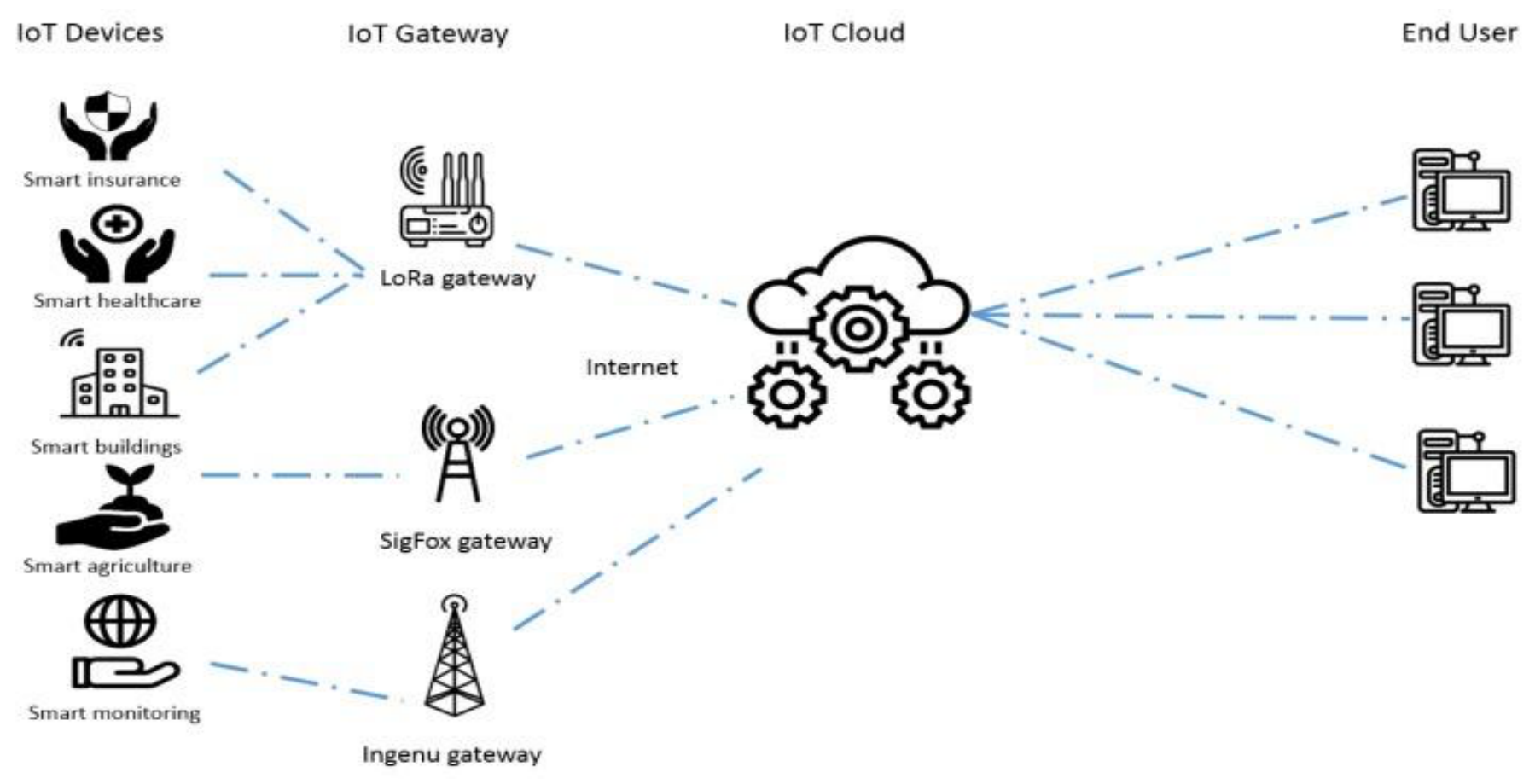

*Fig. 3: Long-Range IoT*

Figure 3 is an example of long-range IoT [4]. For instance, in such rural and distant places, the CAPEX and OPEX expenses of developing and maintaining a terrestrial communication infrastructure are not balanced against the financial benefits. Among other 5G application areas, strategies to expand Internet access to smart farming and smart communities must surpass the farming infrastructure [5]. If they have a cyclical payload or a translucent payload, satellites can function as 5G-gNBs or as RNs connecting 5G UE and 5G RAN. The satellite gateways, which can function as 5G-gNBs or connect to 5G-gNBs through terrestrial links, could be taken into account in the same ways which include in the aerial sector 5G gNBs and 5G CN [4]. IoT gadgets might have a 5G UE built in. The disadvantage of IoT devices is that they are typically extremely basic things with very stringent size, technology, and resource limitations. This is why having 5G UE capabilities could be too much of a strain.

### *D. UAV Architecture*

The 5G wireless communication is planned to incorporate a number of resources to support heavy traffic and a number of services. The convergence of processing, networking, and caching capacities will define this. In addition to acting as an edge computing platform for IoT devices that have limited information processing, UAVs can also act as a complementary method to prefetching some famous components for lowering internet backbone exertion and data transfer latency during peak hours. UAV is a major aspect of IoT and the future 5G networks [1].

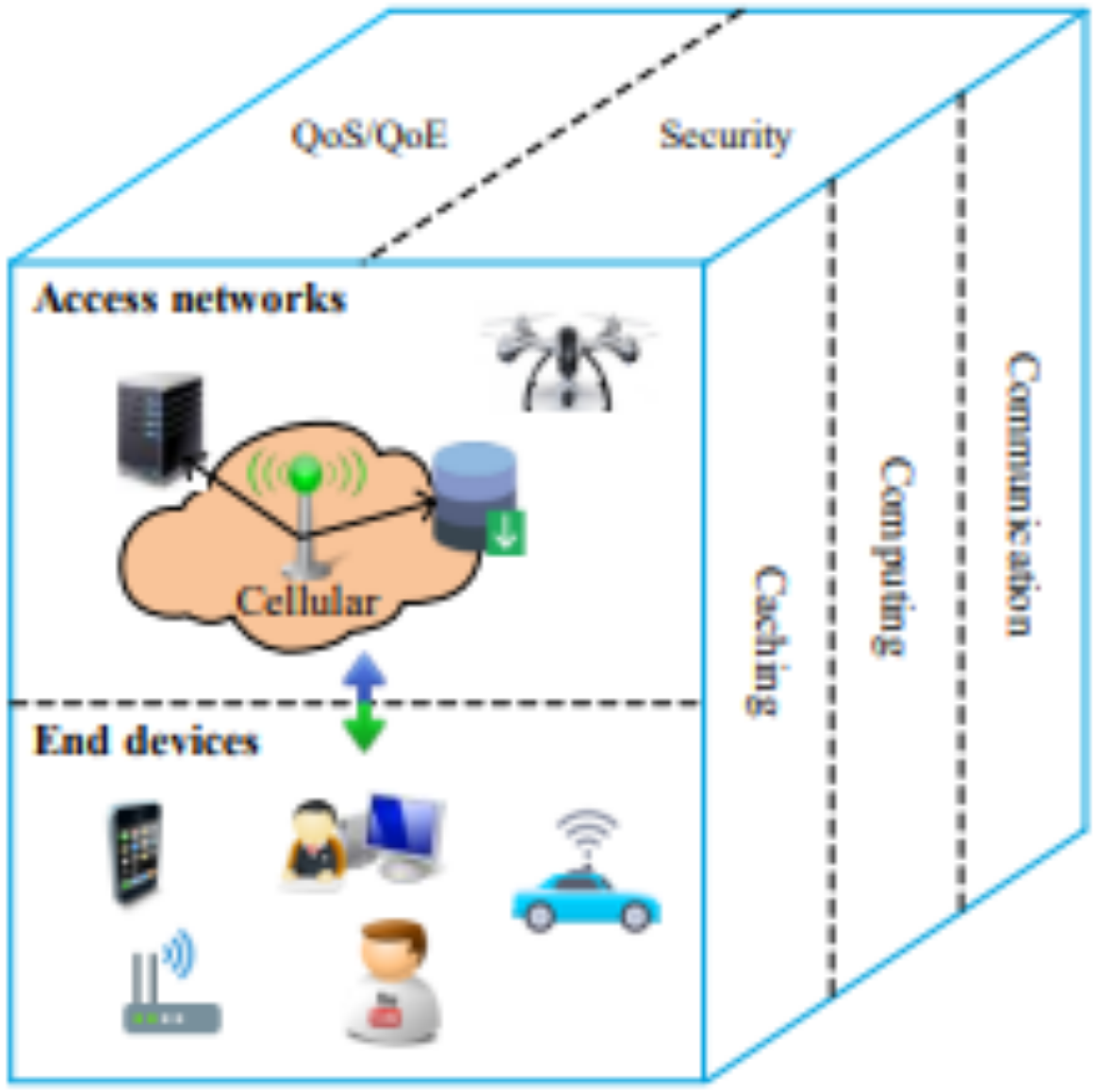

*Fig. 4: The Architecture of UAV and The Computing*

Figure 4 shows the architecture of UAV and the computing behind it. In networks with UAVs, asset mobile devices can transfer their computationally expensive tasks to a hovering UAV having high processing capability and variable connection just at edge nodes, conserving their fuel and lowering the traffic burden at the fixed cloud servers. As a result, the UAV outfitted with a MEC server provides potential benefits in comparison to the traditional terrestrial cell network using fixed BSs. According to this model, a fleet of UAVs can cooperate with one another to avoid collisions and consistently interact with one another while using the mobile spectrum alongside terrestrial customers. The integrated processor allows the handheld devices to independently carry out their own duties but doing so will use up a lot of their local processing power and resources. In this case, the MEC server founder in UAVs will conduct the calculation work in place of the mobile devices after the mobile devices have been given permission to transfer their intense computing tasks to it immediately. Every mobile device was connected directly to a near UAV node that has ample battery capacity and processing capacity right now. [1].

### *E. MEC & UAV*

A technology known as Mobile Edge Computer (MEC) pushes computing and communication capabilities closer to the end user to the edge of a network. For applications that demand real-time processing or low latency, such as augmented reality, virtual reality, and driverless vehicles, this enables faster processing and lower latency. By giving the UAV MEC capabilities on board, MEC can be applied in Unmanned Aerial Vehicles (UAVs) in one way. As a result, there would be less need for the UAV to transfer substantial volumes of data back and forth to the network's central hub. This not only lowers latency but also lessens network load and boosts speed in general. A UAV with MEC capabilities, for instance, might be utilized for surveillance or inspection duties. The MEC capabilities on board would enable the UAV to process this data in real time. The UAV may be outfitted with cameras and other sensors to acquire information about its surroundings. This would eliminate the need for the UAV to transmit all of its data back to a centralized point for processing, allowing it to make decisions and conduct actions depending on the data it is gathering. UAVs may have onboard MEC capabilities in addition to being able to connect to MEC nodes at the network's edge. In doing so, the UAV would be able to access the greater processing power and storage capability of the MEC node, increasing its ability to manage more data and carry out more difficult tasks. MEC-enabled UAVs have the ability to completely transform a variety of markets and uses. They could be utilized for a variety of functions, including delivery, inspection, surveillance, and more. MEC-enabled UAVs have the ability to enhance performance, decrease latency, and enable novel use cases that weren't previously achievable by extending computation and communication capabilities to the edge of the network and enabling the operation of unmanned aircraft. It is difficult for IoT devices to run practical applications because of their constrained battery life and poor computing power. Thankfully, MEC has lately become a viable model to address this problem. By enabling cloud computing capabilities, phone devices can offload their calculation activities toward the edge nodes with the implementation of a MEC server.

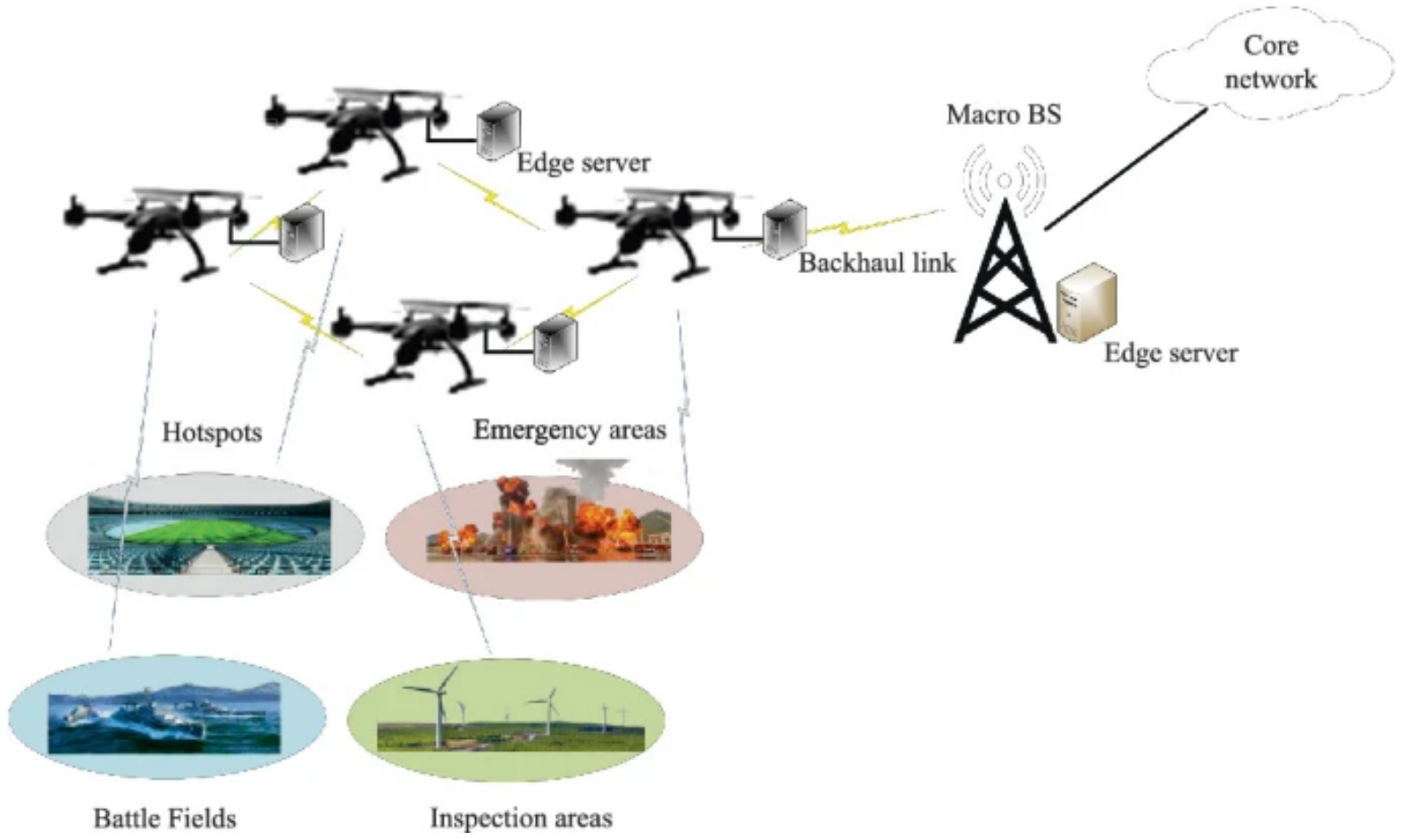

***Fig. 5: Implementation of a MEC Server***

In an emergency situation, land MEC facilities could've been damaged in such a catastrophe, making it impossible to calculate or carry out many rescue actions. To calculate the recovery duties in time, UAVs equipped with edge servers may be sent. Due to a hostile environment, it can be challenging to set up land MEC networks to calculate design evaluation, for example for blades at offshore wind parks or electric lines inside a smart grid. A UAV-assisted MEC network could be extremely useful in this situation. Hotspots can encounter a significant amount of computing activities being outputted by thousands of smartphone users that can deplete the computation capabilities of edge servers, increasing processing latency and lowering user experience [6]. This is what brings in the program delay decrease if a remote endpoint has a lot of computing power and because the program has been run by a remote computer, battery life was improved. These asset mobile devices in UAV-enabled systems can transfer their computationally expensive duties to such a hovering UAV with strong computing capability and variable connection at the end of the network, conserving their power and lowering congestion burden just at the stationary cloud server. As a result, the MEC server-equipped UAV offers prospective benefits over the traditional land cellular networks using static BSs. An integrated processor enables mobile devices to independently carry out their duties but doing so will use a significant portion of available limited processing resources and power. Therefore, the MEC server co-located in UAVs would conduct the calculation work on behalf of the smartphones after the smartphones have been given permission to transfer their intense processing to it directly. Every smartphone is connected directly to a neighboring UAV unit which has ample life of battery and processing capacity right today.

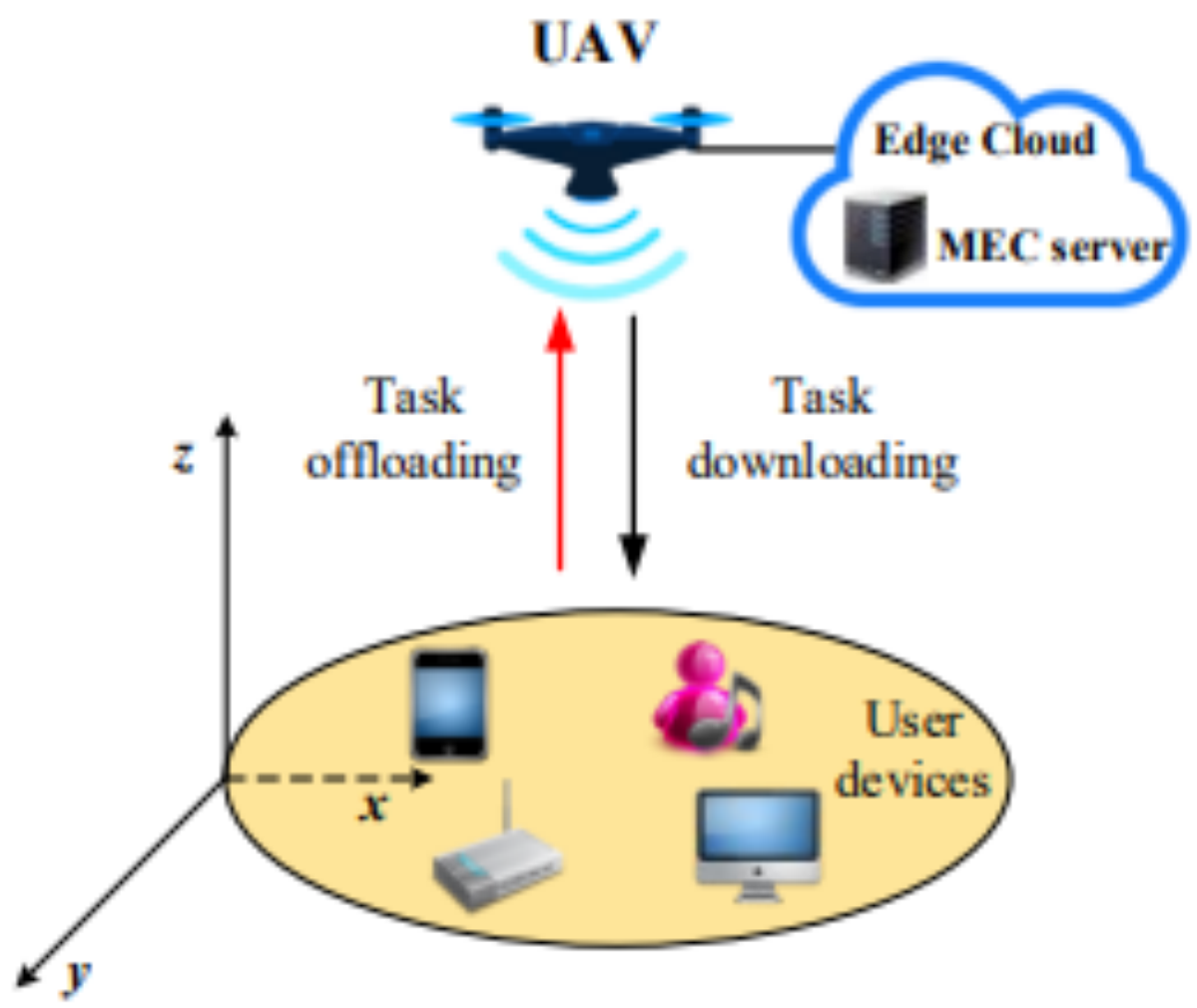

***Fig. 6: MEC Offloading/Downloading***

The above figure shows an example of a UAV and a MEC system This gives user devices just on the network the chance to unload applications' were used to swiftly build an interwoven unloading infrastructure for a fluid online community powered by fog nodes installed on the UAVs. It was possible to distribute compute and communication strain from the cloud layer to an edge network using the distributed UAVs with high-speed computing packets, thereby lessening the stress just on data centers and the traffic volume. Compute lifting and path planning together in order to minimize energy usage in a UAV-enabled MEC system. Within the scenario under consideration, the UAV served a dual function, serving as both an originator and a device that offloaded compute chores from portable devices. In contrast, a UAV's onboard computer could not provide sufficient processing power because of the compact size of such a UAV. Configuration files cannot be executed efficiently as a result. By utilizing calculation shifting to distant servers or adjacent border servers through MBS or SBSs, the energy and calculation efficiency of UAVs can theoretically be improved. As a result, a UAV node does have the choice of executing a program to use its available resources or sending the computation to just a network edge or a distant cloud for processing in accordance with the QoS needs of the application.

*F. IoT Devices*

In order to guarantee that cell membrane LPWA technologies are able to provide effective secure communication again for various scenarios both over Massive IoT and Critical IoT, the below figure lists some of the crucial requirements needed when thinking about the large-scale implementation of such services. These requirements include lower installation cost, long battery life, low system cost, extended transmission range, assistance again for vast quantities of smart devices, and security [7].

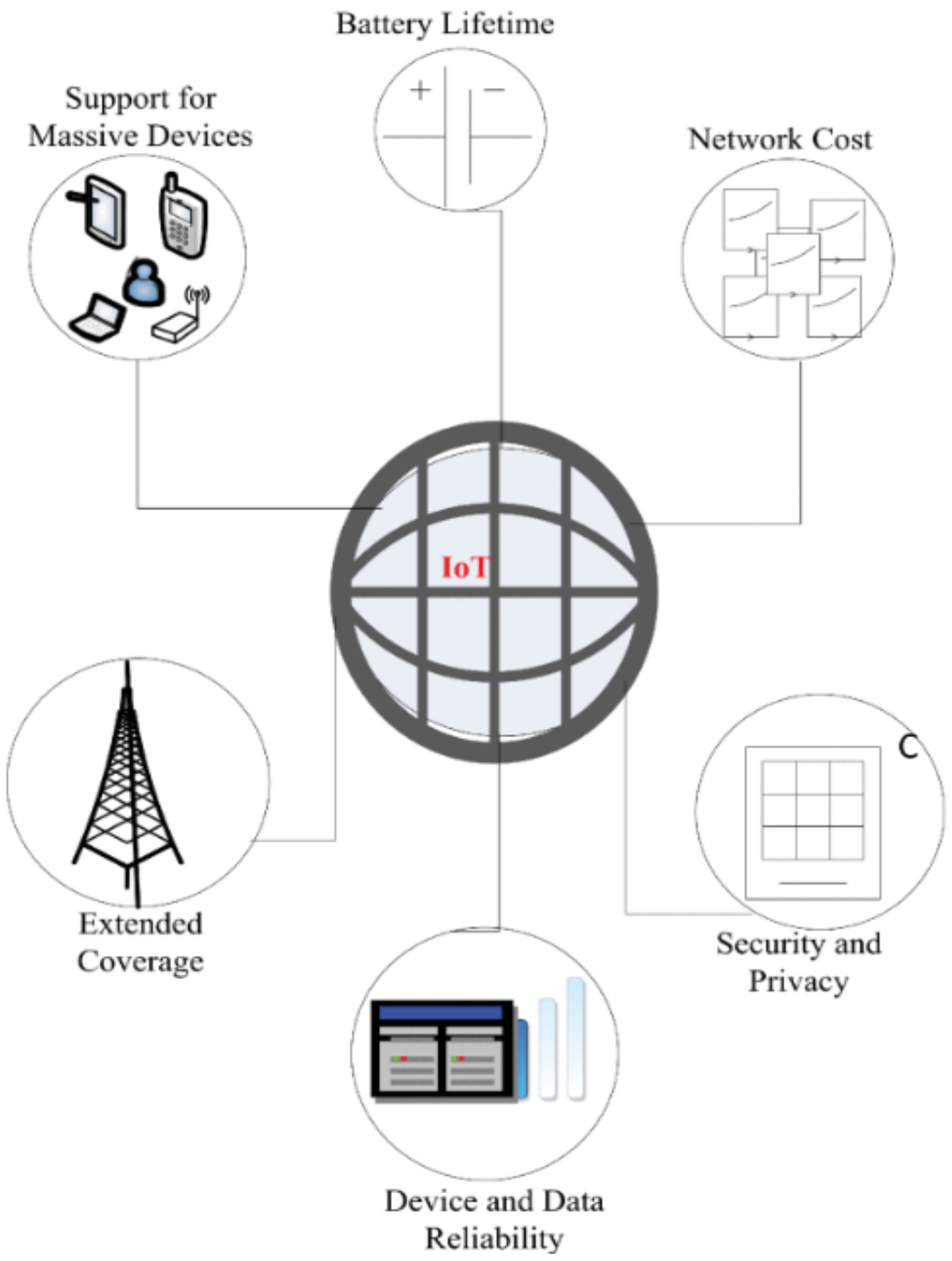

***Fig. 7: IoT Scope***

One essential asset that an IoT can have is strong battery life. Given that almost all IoT devices are rechargeable batteries and are anticipated to operate for an extended amount of time without human involvement, energy efficiency is possibly the most crucial feature of IoT. The software and hardware must take energy efficiency into account when being designed. Many medium access control (MAC) protocols provide job cycling, which enables the transmitter to be placed to sleep for intervals even when it's not anticipating receiving data. This increases battery life. By the use of illumination procedures and schedule efficiency, energy management strategies are also crucial for low power mode [8]. Internet of Things (IoT) devices are becoming more and more prevalent in our daily lives, and their increasing connectivity has both benefits and drawbacks. The capacity of IoT devices to automate numerous operations and processes is one of their key benefits. Smart appliances may be managed remotely and can even buy their own new components when they break down. For instance, smart thermostats can learn our cooling and heating preferences and alter the temperature appropriately. These talents have the potential to not only improve the convenience of our lives but also to reduce time and expense. IoT device connectivity has increased, however, this also brings up privacy and security issues. The risk of data breaches and cyber-attacks grows as more of our personal information is gathered and kept by these gadgets. Users must be aware of the hazards and take precautions to secure their personal information, and manufacturers must emphasize the security of their products. This can involve employing secure passwords, keeping the device's software up to date, and exercising caution while sharing information on these devices. IoT device problems include the environmental impact in addition to security and privacy issues. The persistent internet connection needed by many of these devices can result in higher energy usage. Manufacturers must think about how their products will affect the environment, and consumers must be aware of how much energy their IoT gadgets are using. Utilizing energy-efficient technology and shutting off appliances when not in use are two examples of how to do this. The growing interconnectedness of IoT devices has the potential to alter how we interact with our surroundings. Smart cities, for instance, collect information about traffic patterns, air quality, and other elements of city life using sensors and other technology. This information may then be utilized to increase the efficiency and sustainability of urban areas. Similar to other industries, agriculture can benefit from the deployment of IoT devices to monitor crop health and improve irrigation systems, resulting in more effective and sustainable agricultural methods. The likelihood of a lack of uniformity is one potential problem with the growth of IoT devices. It can be challenging for these devices to interact with one another, and function flawlessly together given the wide variety of products and manufacturers available today. Users may become frustrated as a result, and the full potential of IoT technology may be diminished. Additionally, the ethical implications of how IoT usage balances convenience and surveillance are brought up.

Many of these gadgets make our lives easier and provide valuable capabilities, but they also have the ability to track our whereabouts and collect personal data. Manufacturers must be open and honest about the data they gather and how they use it, and consumers must be aware of the possibility of monitoring and making wise choices about the gadgets they use. The utilization of IoT devices has the potential to develop new industries and job opportunities. There will be a demand for experts in fields like cybersecurity, data analysis, and IoT device creation as more and more devices become connected. These brand-new employment possibilities may encourage economic expansion and innovation. The risks and potential negative effects of IoT technologies must be carefully considered, even though they have the potential to significantly better our lives and make tasks more convenient. We can fully enjoy the advantages of IoT devices while simultaneously guaranteeing the security and privacy of our personal information by being aware of the potential negatives and taking efforts to avoid these risks. In summary, the Internet of Things has the potential to drastically alter how we live and work, but it is crucial to carefully weigh the dangers and potential drawbacks of new technologies. We can fully enjoy the advantages of IoT devices while simultaneously guaranteeing the security and privacy of our personal information by being aware of the potential negatives and taking efforts to avoid these risks. The effect of an amount on IoTDs just on PPO algorithm in comparison to the baseline policies is shown in Fig. 8. It should be emphasized that because each IoTD has more frequent scheduling, the suggested algorithm can reduce the ESA for fewer IoTDs. However, we can observe that the ESA rises for a large number of IoTDs because more scheduling is required to reduce the ESA. Additionally, since the hovering with greedy policy always chooses an IoTD the with highest AoI value, it is more successful than just the randomness policy. Additionally, we can see that the suggested algorithm performs better than all of the baselines. This is predicted given that the standards aren't able to understand the IoTDs' activation pattern and do not take the UAV's altitude adaptation into account (spectrum.library).

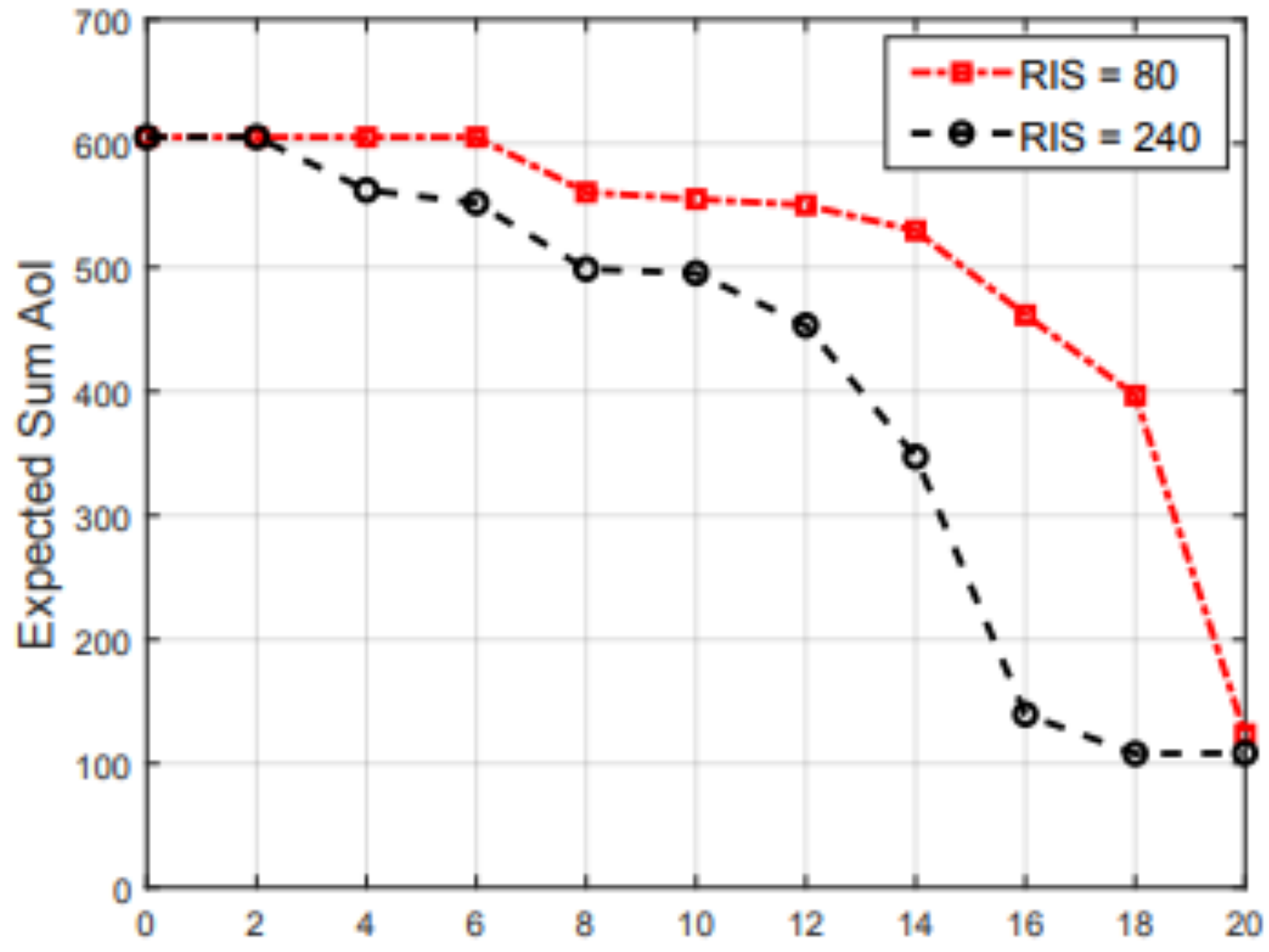

***Fig. 8: Expected Sum AoI***

## III. DISCUSSION

5G is the fifth and most recent generation of mobile networking technology, promising much faster speeds, lower latency, and more capacity over 4G. With various applications from virtual and mixed reality to self-driving cars and the Internet of Things, the launch of 5G has the potential to alter the way we live and work (IoT). One of the most important characteristics of 5G is its improved speed, which enables quicker download and upload rates, as well as more efficient video as well as other data-intensive application streaming. This is accomplished by employing high-frequency bands, such as mm-wave frequencies, which can transport more information while having a smaller range than earlier generations' lower frequency bands. As a consequence, 5G networks will require additional infrastructure to assure coverage, such as more cell phone towers and tiny cells. Another big benefit of 5G is its low response time or the amount of time it takes for data to be delivered and received. This is critical for applications requiring actual dialogue, such as augmented and virtual reality or self-driving automobiles, where even little delays can have major repercussions. 5G has a delay of about 1 millisecond, whereas 4G has a delay of about 50 milliseconds, rendering it far more appropriate for these sorts of applications. In addition to improved speed and latency, 5G provides increased capacity, enabling further devices to also be connected to the network at one time without performance degradation. This is especially essential given that the number of linked devices is predicted to exceed 75 billion through 2025. Whereas the possible advantages of 5G are substantial, there are some worries about its deployment. The possible impact on health is one source of worry since the higher frequency bands utilized in 5G have been connected to a rise in radiation exposure. However, according to the World Health Organization (WHO), the amounts of radiation released by 5G technology are all within international norms and do not represent a substantial health risk. Another source of worry is the expense of deploying 5G networks, even as the necessary infrastructure is substantially more complicated and costly than in earlier generations. This has resulted in competition among nations and firms to build 5G networks, despite some worries about security and espionage. 5G has the ability to boost economic growth and offer new job possibilities in addition to its potential uses. The deployment of 5G networks is predicted to need major infrastructure investments and technology, resulting in the development of new employment in a range of industries such as engineering, construction, and information technology. However, deploying 5G networks isn't without difficulties. Aside from the expense and difficulty of constructing the network, there are also worries regarding cybersecurity and the possibility of spying. As more gadgets connect to the internet, their potential for cyber assaults rises, and governments and businesses must take precautions to safeguard networks and preserve critical data. Despite these obstacles, 5G has considerable potential advantages and is projected to have a big influence on a wide range of sectors and applications. [11] As more nations and businesses deploy 5G networks, as well as technology, become more generally available, substantial breakthroughs in industries including transportation, health, and information are inevitable. 5G has other features that make it an appealing technology for both consumers and enterprises, in addition to higher speeds, lower latency, and more capacity. One of these advantages is the capacity to support increased density or the amount of devices which can connect to a network in a given space. This is especially beneficial in congested metropolitan areas where the connection is in high demand. Another advantage of 5G is that it can enable novel use cases including massive machine-type communications (mMTC) and ultra-reliable low-latency communications (URLLC). The capacity to link a large number of devices with modest data needed, including such sensors and other IoT devices, is referred to as mMTC. The capacity to deliver very low latency and great dependability for crucial applications such as self-driving vehicles and remote surgery is referred to as URLLC.

5G also has better energy efficiency than previous generations, which is significant as the demand for mobile data grows. This is accomplished through a mix of strategies, including the use of sophisticated modulation schemes as well as the ability to turn off certain network components when they are not being used. Furthermore, 5G is intended to support a variety of frequency bands, including low and high frequencies, allowing it to function in a wide range of situations and give coverage both in urban and rural locations. Overall, 5G represents a significant advancement in mobile networking technology, providing a variety of benefits that have the potential to alter the way we live and work. While there are still challenges to overcome, such as the cost and complexity of constructing the necessary infrastructure, as well as concerns about cybersecurity, the potential benefits are substantial and are expected to propel economic growth and generate new jobs in a wide range of industries. Overall, 5G represents a significant advancement in mobile network technologies, providing a variety of advantages that have the potential to alter the way we live and work. Although there are still challenges to overcome, including the cost and complexity of constructing the necessary infrastructure, as well as worries about cybersecurity, the possible advantages are substantial and are expected to propel economic growth and generate new jobs in a wide range of industries. The 5G architecture's fundamental purpose is to support wireless device connectivity. These devices presently consist of smartphones and/or tablets, but in the future, they will also comprise a variety of other items, such as automobiles and medical technologies. The 5G Network Architecture, see Fig. 9, includes two main components, as shown in the picture, to aid with its breakdown. Its utilization of the Mobile Core, which has a variety of functions, is its first notable feature. Mobile core assists in tracking the user to ensure that there is no service disruption and that they are being paid and charged appropriately. Additionally, it offers IP correspondence for mobile data. The Radio Access Network is the second component (RAN). The radio access network monitors the radio spectrum and makes sure that the end user receives the stated level of quality [12].

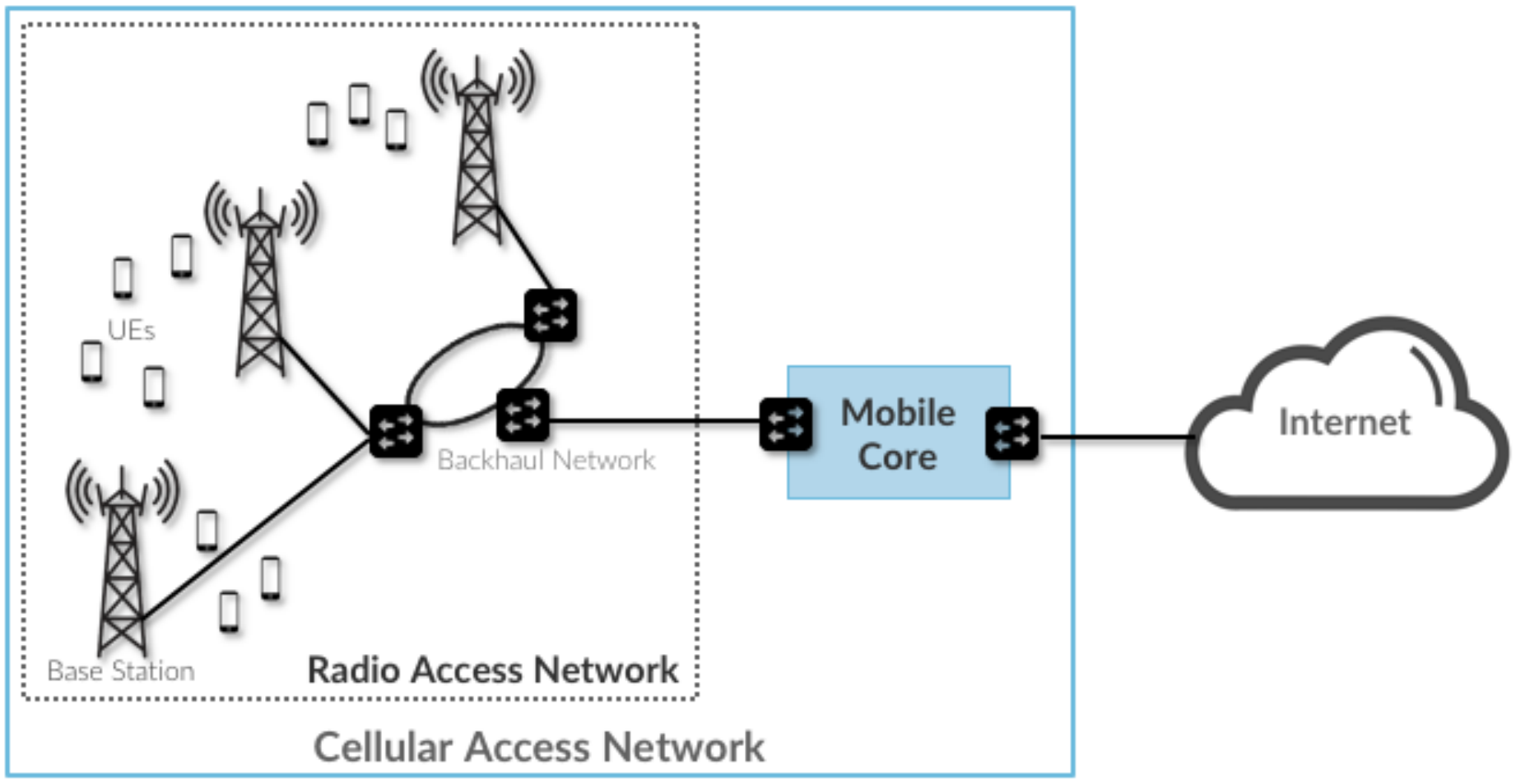

***Fig. 9: Basic 5G Architecture.***

5G is the fifth and most recent generation of mobile networking technology, promising much faster speeds, lower latency, and more capacity over 4G. With various applications from virtual and mixed reality to self-driving cars and the Internet of Things, the launch of 5G has the potential to alter the way we live and work (IoT).

### *A. 5G SECURITY IDENTIFICATION*

One of the main harmful actions that aid in the penetration of a system is a denial of service (DoS) assault. The first, flooding, and the second, semantic DoS attacks are the two types of DoS that we need to be conscious of and develop prevention strategies for. Furthermore, the analysis in the section that follows shows that SBCA has a far higher query capacity than conventional centralized systems. The next is DoS Flood Assault Prevention where SBCA decentralizes authentication operations from the AUSF/ARPF to everyone gNBs to mitigate the effects of legal request flooding. The combined computing capacity of more and more gNBs makes it harder for enemies to overpower and seize the entire 5G network. Compared to traditional centralized systems, SBCA is significantly better at processing demands. As a result, listeners are unable to recognize a specific feature or errors that caused signals to find a device. With SBCA, the activation key is created using randomly generated public ECDH keys. As a result, even if all future temporary keys are stolen, attackers will not be able to recover the session key. Its confidentiality won't be compromised even if the session key is.

### *B. IRS AND UAV*

M-MIMO and millimeter wave communications have some intriguing advantages, but their high complexity requirements, high hardware costs, and higher energy usage are still major obstacles to their widespread use. A novel and affordable alternative to boost received power and reduce wind disturbance over 3D space are IRS, which has recently come into existence. An IRS is made up of numerous passive reflecting components, each of which has an adjustable reflection coefficient and is capable of reflecting the electromagnetic wave that is impinging on it. An IRS is able to modify the communication links so that the desired signals are added cogently, and interference is forced to cancel at designated receivers, significantly increasing the interaction throughput without the requirement for deployment. This is done by intelligently trying to coordinate the observations of all elements [9]. Additionally, IRSs have practical features that make them appealing, such as their small weight, which makes it simple to place them on surfaces like walls or even the moving surfaces of rapidly flowing vehicles to support a variety of applications. Therefore, IRS has been viewed as a game-changing technique for turning our present wireless atmosphere into a smart one, which might have positive impacts on a broad variety of industry verticals like transport, production, and smart cities. Additionally, recently acknowledged as just a groundbreaking technology for the upcoming sixth-generation 6G ecosystem is IRS. In principle, IRS can be placed on the ground to support UAV communication or connected to UAVs to support terrestrial communications, as explained below. This is illustrated in Figure 10 moreover, a quick comparison of previous IRS works [9].

].

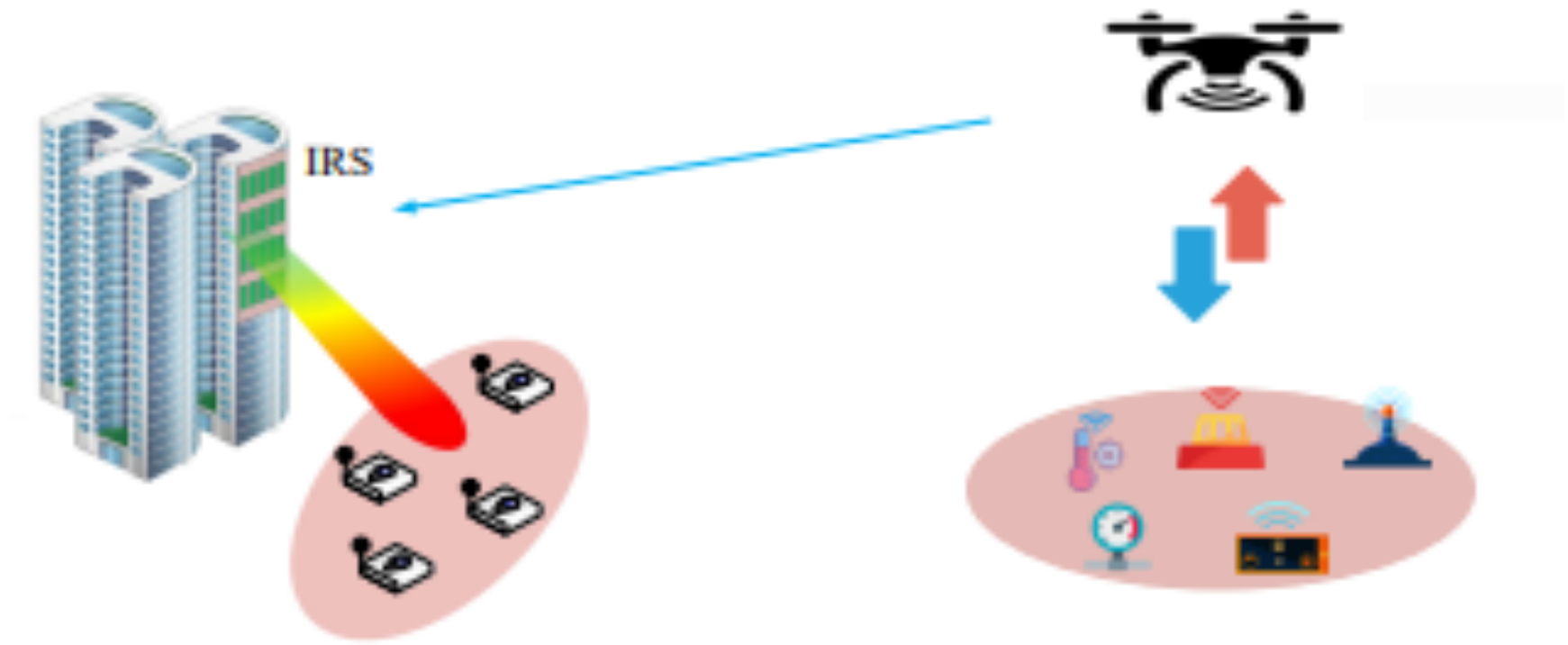

*Fig. 10: UAV Physically Monitoring IoT Devices*

A UAV just has to physically monitor a portion of IoT devices, as indicated in Fig. 10, and the large route lost here between the UAV as well as other devices may be successfully compensated by utilizing an installation of IRS. Due to the restricted power generation of IoT devices, this is especially helpful for uplink broadcasts as it helps balance the trade-off in surface connections. Additionally, as was already indicated, existing cellular networks often have downgraded ground BS antennas with primary lobes that point downward to level components for the base station and only support side lobes for UAVs flying above the BSs. In fact, recent 3GPP tests have shown that UAVs may complement current terrestrial BSs by picking up their patchy signals from them [9]. By jointly maximizing aerial IRS placement, IRS aspect changes, and transmit beamforming at the BS, the minimum SNR within a certain square service area was increased. The best horizontal location was demonstrated to depend only on the ratio between both the UIRS elevation and the original length, and the authors then determined an optimum UIRS location and phase changes for the unique situation of single-location SNR maximization. Inside the generalized form of area coverage augmentation, a two-development approach was suggested by decoupling the step enhancement from the IRS location layout, in which a 3D beam propagation and straightening technique has been suggested to gain back 3D beam patterns that matched the size of the service area on the ground [9].

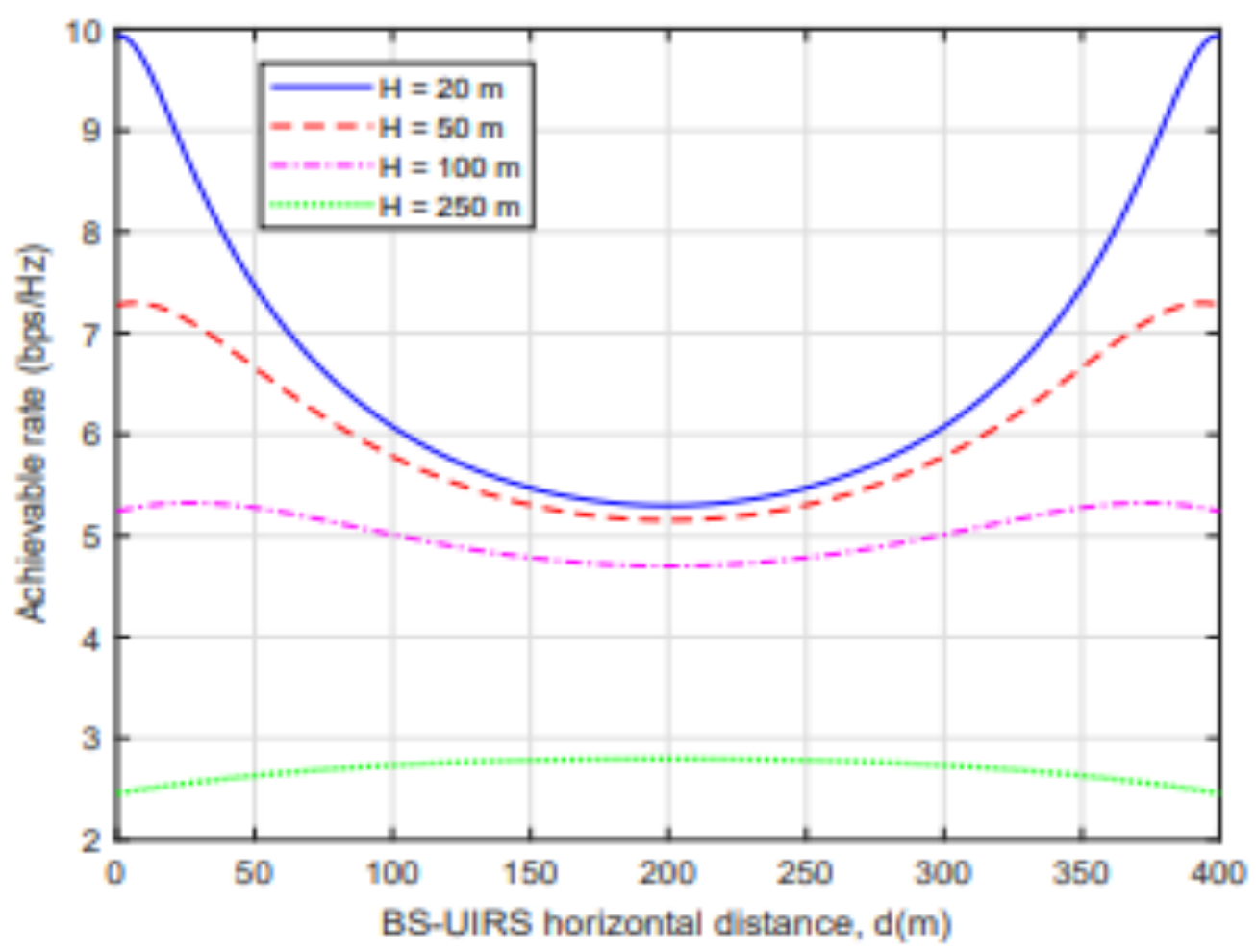

***Fig. 11: Plotting the Attainable Data Rate of The Under-Consideration***

The key benefit of this strategy is that it doesn't require previous information about the changing situation; instead, it picks up on its characteristics as the UIRS provides service, based on the measurements and input it receives at each communication stage. The results demonstrated that by adopting one such UIRS as opposed to a static IRS, considerable benefits can be made in terms of the overall bit rate in addition to the chance of LoS. However, in order to make the system design simpler, it was believed that the BS only transmits and that the UIRS solely reflects data when the aircraft is hovering. For various BS transmit powers, we plot the attainable data rate of the under-consideration UAV-assisted relaying system against the position of the UAV in Figure 11. The total delay is 1 ms, and each mini-time slot has a bandwidth of 200 kHz and TBlock = 0.01 ms. So, the number of blocks that can be transmitted is capped at 100 [9].

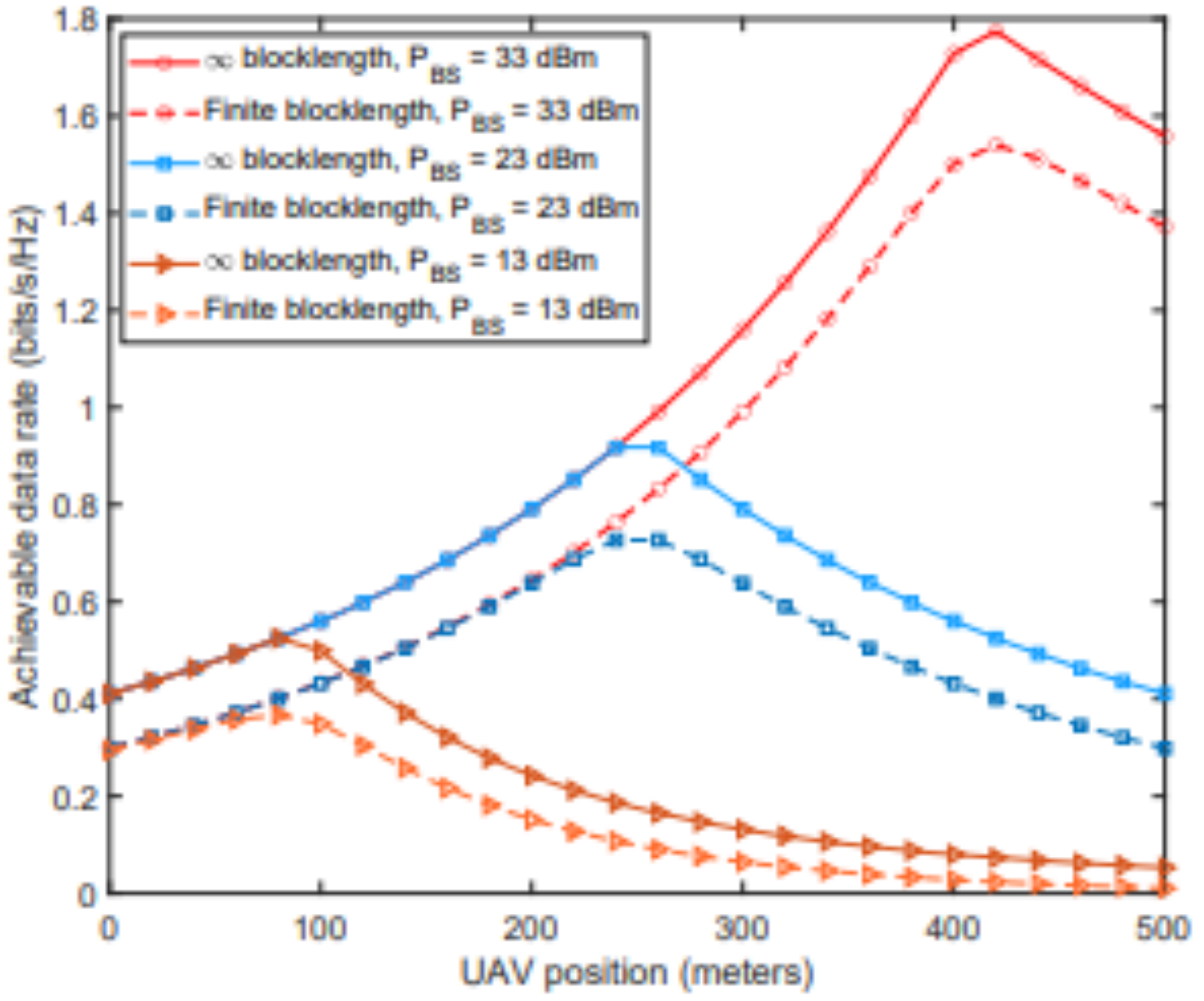

***Fig. 12: Data Rate That a UAV-Assisted Relaying Network***

Figure 12 compares the data rate that a UAV-assisted relaying network can achieve for various BS transmit powers and various block lengths. The UAV's transmission power is 23 dBm while it is flying at a constant height of H = 100 m and there is a base user 500 meters distant from the BS. As seen in Figure 12, there is an efficiency difference between the situations of limited block length and unlimited block length because the former has insufficient coded information blocks for the effect of the Gaussian filter to be adequately averaged out to reach the capacity of the system. Since the edge delay restriction in URLLC is strict, this gap is typically unavoidable [9]. Furthermore, the power budgets between the UAV as well as the grounded BS have a significant impact on the best UAV position, which maximizes the end-to-end attainable rate. In reality, the ideal UAV position makes an effort to balance the system's two hops' achievable rates.

### *C. UAV COMMUNICATION*

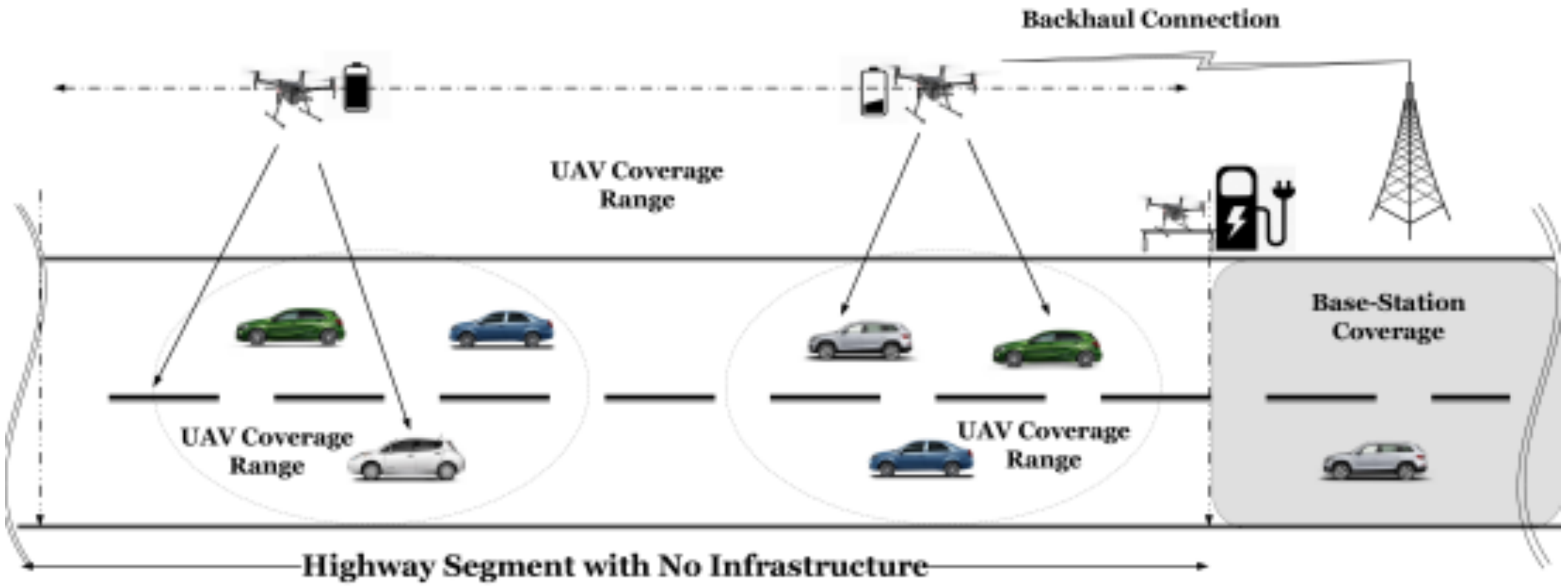

***Fig. 13: Two Hops System***

As depicted in Figure 13, we take into account a section of the highway with communication infrastructure. that is either missing or destroyed (such as after a natural disaster). Additionally, this sector features a one-way flow of traffic for cars leaving the range of a permanent base station (BS). It is believed that the UAVs have large capacity front haul connections with input ground BS, such as available storage optics (FSO) or mm (mmWave) links, where a central unit with an actor-critical agent resides. The Actor-Critical Agent handles the collaboration between the deployed UAVs while continuously learning the best trajectory strategy from observations of the changing vehicular environment. As a result, other deployed UAVs will cover a vehicle that is not covered while it is inside the range of one UAV [10]. This interference-free model has been extensively utilized. assumes that each UAV can simultaneously interact with many cars inside its range by assigning the proper orthogonal resources to achieve interference-free communication. Additionally, we presumptively allot distinct portions of the spectrum to nearby UAVs such that inter-UAV transmission is likewise interference-free. Therefore, make the assumption that the cars are supplied when they are inside a UAV's coverage area and are not being interfered with by other UAVs. As a result, the focus of our study is on how to address the UAVs' coverage issue. Another performance statistic the survey looks at is the proportion of served IoTDs. This statistic is shown in Figure 14 in relation to the service fee and for various dates. With an increase in service level quantity Smin per IoTD, it is obvious that the UAV will require additional time and wireless resources to collect information from one sensor before moving on to collect data from another [10].

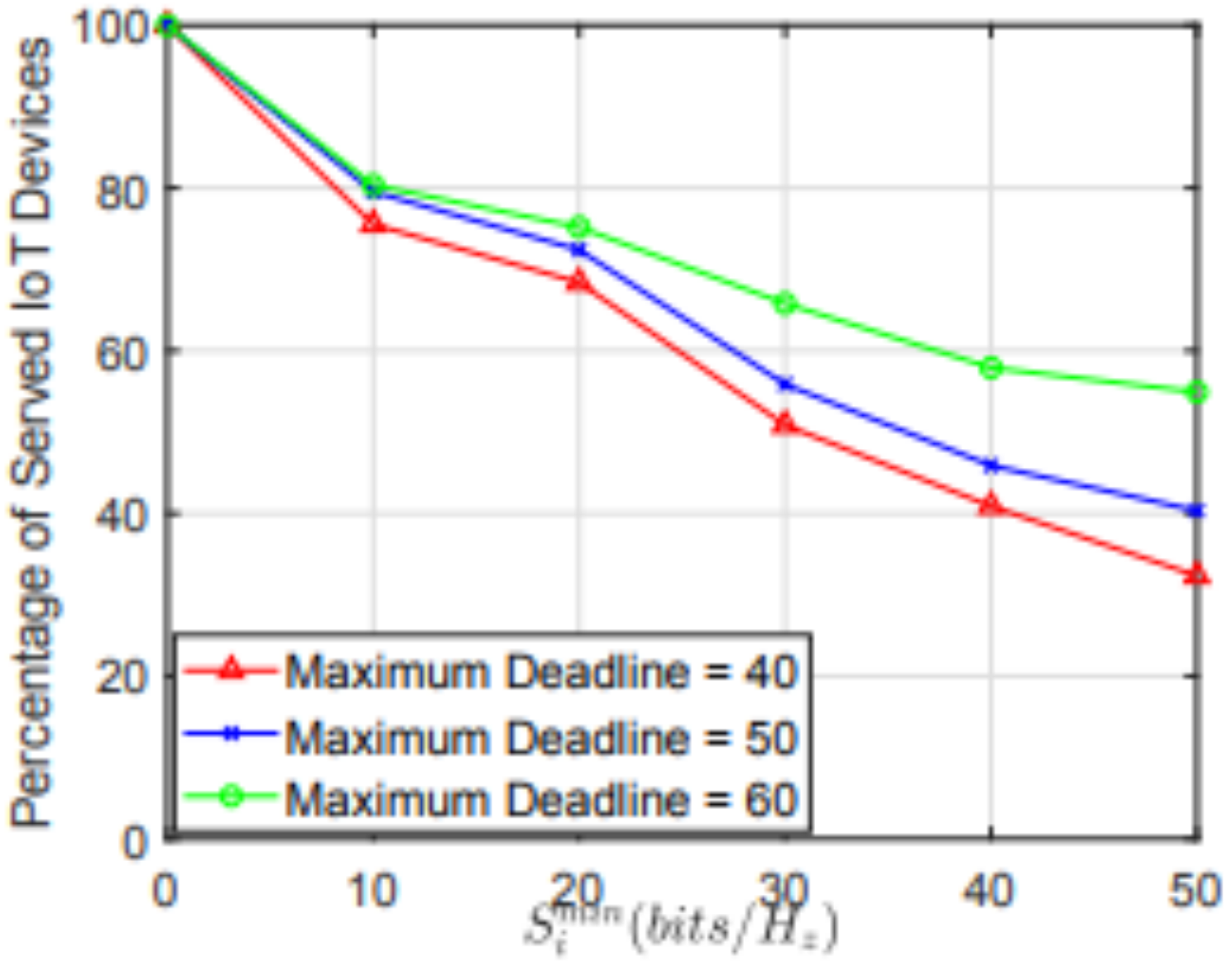

***Fig. 14: Service Level Quantity Smin Per IoTD***

In addition, compared to demands that are more stringent, less harsh constraints will provide the UAV more time and money to serve more devices. In comparison to the network size, which represents the greatest number of IoTDs that can be found in a given area, Figure 15 shows the proportion of served IoTDs. When can be observed, even as the quantity of IoTDs in the same area rises, the percentage of served IoTDs falls. Fewer multiple antennas are allotted for each IoTD when more devices are taken into consideration in the same area since spectrum bands and flying hours are restricted. By extending the deadlines, the percentage of IoTDs that are supplied will rise as anticipated because the UAV will have more time to devote to more resources [10].

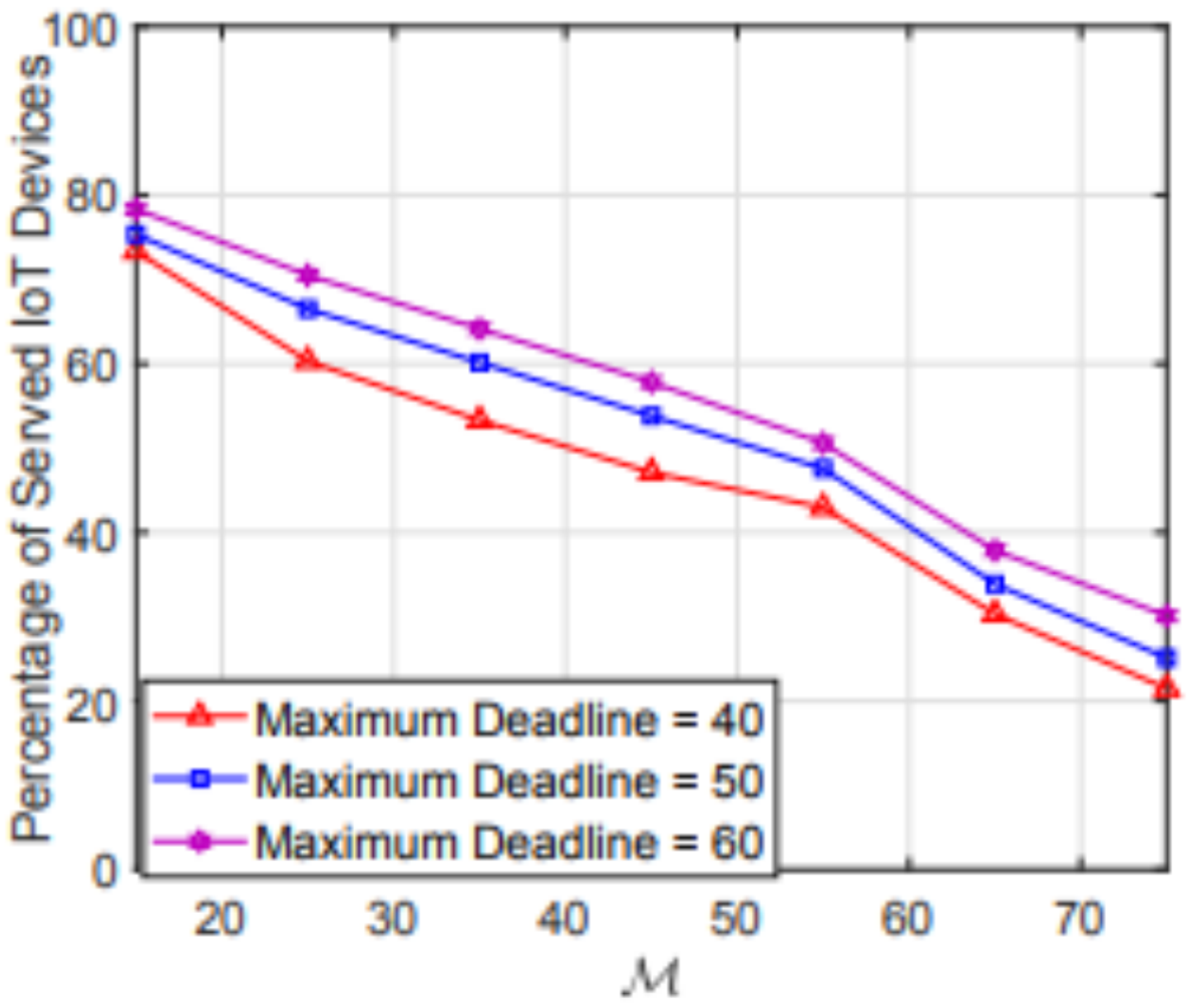

***Fig. 15: Proportion of Served IoTDs***

***D. UAV LAW***

With the growth of the Unmanned aerial vehicles industry, laws and regulations have been put in place to ensure the safe and responsible use of these vehicles. The Federal Aviation Administration (FAA) in the United States is in charge of policing drone use. The FAA has issued "part 107 rules" for drone operators, which specify the conditions for using drones for business reasons. According to these regulations, drone operators must receive a Distant Pilot Certificate, which calls for passing a written exam and proving that they are knowledgeable about drone safety and operation. The FAA has issued regulations for using drones for recreation in conjunction with the part 107 standards. According to these regulations, referred to as the "Special Rule for Model Aircraft," drone operators must fly their craft at altitudes no higher than 400 feet, maintain constant visual contact with them, and operate them at least 5 miles from airports. State and municipal laws may also apply to the use of drones in addition to the FAA's rules. For instance, some jurisdictions have regulations that limit the usage of drones for surveillance or hunting, or that forbid their use over specific properties. When using their drones, drone operators must be informed of and abide by all applicable rules and regulations. Drone use raises privacy issues as well because it may be fitted with cameras and other sensors that could capture and send data. Some states have passed legislation limiting the use of drones for monitoring or requiring drone pilots to get permission before flying over specific properties in response to these worries. There are laws and regulations pertaining to the manufacture and selling of drones in addition to laws and regulations pertaining to the operating of drones. These laws and rules may include criteria for testing and certification and are intended to assure the security and dependability of drones. Some different points to take into consideration when speaking about UAVs and their laws are [13-17].

- **Cybersecurity:** Drones can be subject to cyberattacks, much like other linked gadgets. This raises questions about the possibility of drones being utilized for evil deeds like conducting cyberattacks or data espionage. There are rules and regulations pertaining to drone cybersecurity, including specifications for safe communication and data storage, to address these problems.
- **International laws:** There are laws and rules governing drones outside of the US as well. Drone operators must be aware of and abide by these rules when using their drones outside of their home country because many other nations have their own drone-related laws and regulations. Many nations utilize the International Civil Aviation Organization's (ICAO) drone operating standards as a model for creating their own drone laws and regulations.
- **Liability and insurance:** Liability concerns for the drone operator or pilot may arise in the event of a drone accident or incident. Many laws and regulations oblige drone operators and pilots to obtain insurance to insure against potential losses or injuries in order to allay these worries. Drone operators and pilots must be aware of their insurance and responsibility responsibilities under all applicable laws.

- **Privacy and surveillance:** As were already indicated, privacy and surveillance issues are raised by the deployment of drones that are outfitted with cameras and sensors. There are laws and regulations that could relate to the gathering and use of data obtained by drones in addition to specific laws and regulations governing the utilization of drones for surveillance. Such laws and regulations must be understood and followed by drone operators and pilots.
- **Military and government use of drones:** Drone usage extends beyond just business and entertainment. Drones are also used by the military and government organizations for a variety of operations, including intelligence collection and targeted attacks. Additional laws and restrictions, such as those concerning the use of violence as well as the defense of civilians, frequently apply to certain drone usage.

In conclusion, a complicated network of regulations and laws at the national, regional, and local levels govern the usage of drones. Drone operators must be informed of and abide by all applicable rules and regulations, and drone vendors and manufacturers must make sure that their offerings meet all necessary safety and dependability requirements.

## IV. CONCLUSION

Drones, commonly referred to as unmanned aerial vehicles (UAVs), are aircraft that are flown remotely or automatically without a human pilot present. Their adaptability and affordable price have helped them gain popularity in recent years. Aerial photography, transportation of commodities, rescue and search missions, and military activities are just a few of the many uses for UAVs. UAVs have the potential to change sectors including farming, building, and transit in the future, playing an even bigger role in society. As technology advances and legal structures were put in place to guarantee their safe operation, the use of UAVs will probably continue to grow. As the quantity of IoT mobile devices increases quickly, a significant degree and broadband connectivity communications network that really can stably handle multiple IoT devices is required. The hovering UAVs have lately gained significant scientific interest in order to meet this demand. The convergence of IoT systems, UAVs, and their satellites inside the 5G ecosystem is discussed in this study. The various UAV and IoT use cases, and the function of satellites within IoT was discussed including a short introduction to various IoT technologies. By distinguishing the actions of IoT devices, and UAVs, various architectures were proposed. For B5G telecommunications, an unified network of space, air, and was envisioned. The associated design issues that can substantially aid in understanding this newly added network architecture are explored. The topic of UAVs and 5G implementation is a vastly growing topic that will keep growing throughout the years with the interest of many countries and businesses.

Future work includes adjusting our current cybersecurity framework [18-39] to work for the UAVs by extending the blockchain model and federated learning approaches to address the 5G communication security of the UAVs.

## REFERENCES

1. *This article has been accepted for publication in a future ... - arxiv*. (n.d.). Retrieved December 5, 2022, from https://arxiv.org/pdf/1901.06637.pdf
2. Marchese, M., Moheddine, A., & Patrone, F. (2019, August 26). *IOT and UAV integration in 5G hybrid terrestrial-satellite networks*. Sensors (Basel, Switzerland). Retrieved December 4, 2022, from https://www.ncbi.nlm.nih.gov/pmc/articles/PMC6749430/#B24-sensors-19-03704
3. Alfaia, R. D., Souto, A. V. de F., Cardoso, E. H. S., Araújo, J. P. L. de, & Francês, C. R. L. (2022, June 15). Resource management in 5G Networks assisted by UAV base stations: Machine Learning for Overloaded Macrocell prediction based on users' temporal and Spatial Flow. MDPI. Retrieved December 23, 2022, from https://www.mdpi.com/2504-446X/6/6/145
4. Alshaibani, W. T., Shayea, I., Caglar, R., Din, J., & Daradkeh, Y. I. (2022, August 12). *Mobility management of unmanned aerial vehicles in ultra-dense heterogeneous networks*. Sensors (Basel, Switzerland). Retrieved December 22, 2022, from https://www.ncbi.nlm.nih.gov/pmc/articles/PMC9416608/
5. *A survey on 5G networks for the internet of things ... - IEEE xplore*. (n.d.). Retrieved December 5, 2022, from https://ieeexplore.ieee.org/document/8141874/
6. Zhang, Y. (1970, January 1). Mobile edge computing for uavs. SpringerLink. Retrieved December 20, 2022, from https://link.springer.com/chapter/10.1007/978-3-030-83944-4_6
7. *IEEE Xplore*. (n.d.). Retrieved December 23, 2022, from https://ieeexplore.ieee.org/Xplore/dynhome.jsp

8. *Low-altitude unmanned aerial vehicles-based internet of ... - IEEE xplore*. (n.d.). Retrieved December 5, 2022, from https://ieeexplore.ieee.org/abstract/document/7572034/
9. *Escholarship*. eScholarship, University of California. (n.d.). Retrieved December 22, 2022, from https://escholarship.org/
10. *Unmanned aerial vehicles for 5G and beyond: Optimization and deep ...* (n.d.). Retrieved December 23, 2022, from https://spectrum.library.concordia.ca/id/eprint/988009/
11. *Chapter 3: Basic architecture*. Chapter 3: Basic Architecture - 5G Mobile Networks: A Systems Approach Version 1.1-dev documentation. (n.d.). Retrieved December 4, 2022, from https://5g.systemsapproach.org/arch.html#ra
12. *A survey on cellular-connected uavs: Design challenges, enabling 5G/B5G Innovations, and experimental advancements*. Computer Networks. Retrieved December 4, 2022, from https://www.sciencedirect.com/science/article/abs/pii/S1389128620311324?fr=RR-2&ref=pdf_download&rr=773eaaf6bae1c422
13. Drone laws in the U.S.A.: UAV coach (2023). UAV Coach. (2022, December 16). Retrieved December 20, 2022, from https://uavcoach.com/drone-laws-in-united-states-of-america/
14. Flaherty, N. (2020, November 19). *Security risks of drones in 5G Networks*. EENewsEurope. Retrieved December 4, 2022, from https://www.eenewseurope.com/en/security-risks-of-drones-in-5g-networks/
15. *How drones aid in natural disaster response - 42west*. (n.d.). Retrieved December 5, 2022, from https://www.adorama.com/alc/drones-natural-disaster-response/
16. *Integrating drones and wireless power transfer into beyond 5G networks ...* (n.d.). Retrieved December 23, 2022, from http://users.cecs.anu.edu.au/Salman.Durrani/_PhD/Xiaohui_thesis.pdf
17. Unmanned Aircraft Systems (UAS). Unmanned Aircraft Systems (UAS) | Federal Aviation Administration. (n.d.). Retrieved December 20, 2022, from https://www.faa.gov/uas
18. Hisham A. Kholidy, “Multi-Layer Attack Graph Analysis in the 5G Edge Network Using a Dynamic Hexagonal Fuzzy Method”,. Sensors 2022, 22, 9.
19. Hisham A. Kholidy, A. Karam, J. L. Sidoran, M. A. Rahman, "5G Core Security in Edge Networks: A Vulnerability Assessment Approach", the 26th IEEE Symposium on Computers and CommunicationsGreece, September 5-8, 2021. https://ieeexplore.ieee.org/document/9631531
20. Hisham A. Kholidy, “A Triangular Fuzzy based Multicriteria Decision Making Approach for Assessing Security Risks in 5G Networks”, December 2021, {2112.13072}, arXiv.
21. Kholidy, H.A., Fabrizio Baiardi, "CIDS: A framework for Intrusion Detection in Cloud Systems", in the 9th Int. Conf. on Information Technology: New Generations ITNG 2012, April 16-18, Las Vegas, Nevada, USA. http://www.di.unipi.it/~hkholidy/projects/cids/
22. Kholidy, H.A. (2020), "Autonomous mitigation of cyber risks in the Cyber–Physical Systems", doi:10.1016/j.future.2020.09.002, Future Generation Computer Systems,Volume 115, 2021, Pages 171-187, ISSN 0167-739X, https://doi.org/10.1016/j.future.2020.09.002.
23. Hisham A. Kholidy, Abdelkarim Erradi, Sherif Abdelwahed, Fabrizio Baiardi, "A risk mitigation approach for autonomous cloud intrusion response system", Computing Journal, Springer, June 2016. (Impact factor: 2.220). https://link.springer.com/article/10.1007/s00607-016-0495-8

24. Hisham A. Kholidy, “Detecting impersonation attacks in cloud computing environments using a centric user profiling approach”, Future Generation Computer Systems, Vol 115, 17, December 13, 2020, ISSN 0167-739X.
25. Kholidy, H.A., Baiardi, F., Hariri, S., et al.: “A hierarchical cloud intrusion detection system: design and evaluation”, Int. J. Cloud Comput., Serv. Archit., 2012, 2, pp. 1–24.
26. Kholidy, H.A., “Detecting impersonation attacks in cloud computing environments using a centric user profiling approach”, Future Generation Computer Systems, Volume 115, issue 17, December 13, 2020, Pages 171-187, ISSN 0167-739X, https://doi.org/10.1016/j.future.2020.12.
27. Kholidy, Hisham A.: 'Correlation-based sequence alignment models for detecting masquerades in cloud computing', IET Information Security, 2020, 14, (1), p.39-50.
28. Kholidy, H.A., Abdelkarim Erradi, “A Cost-Aware Model for Risk Mitigation in Cloud Computing SystemsSuccessful accepted in 12th ACS/IEEE International Conference on Computer Systems and Applications (AICCSA), Marrakech, Morocco, November, 2015.
29. Kholidy, H.A., Ali T., Stefano I., et al, “Attacks Detection in SCADA Systems Using an Improved Non-Nested Generalized Exemplars Algorithm", the 12th IEEE Int. Conference on Computer Engineering and Systems, December 19-20, 2017.
30. Qian Chen, Kholidy, H.A., Sherif Abdelwahed, John Hamilton, "Towards Realizing a Distributed Event and Intrusion Detection System", the Int. Conf. on Future Network Systems and Security, Florida, USA, Aug 2017.
31. Hisham A. Kholidy, A. Erradi, Sherif Abdelwahed, Abdulrahman Azab, “A Finite State Hidden Markov Model for Predicting Multistage Attacks in Cloud Systems", in the 12th IEEE Int. Conf. on Dependable, Autonomic and Secure Computing, China, August 2014.
32. Ferrucci, R., & Kholidy, H. A. (2020, May). A Wireless Intrusion Detection for the Next Generation (5G) Networks”, Master’s Thesis, SUNY poly.
33. Rahman, A., Mahmud, M., Iqbal, T., Saraireh, L., Kholidy, H., et. al. (2022). Network anomaly detection in 5G networks. Mathematical Modelling of Engineering Problems, Vol. 9, No. 2, pp. 397-404. https://doi.org/10.18280/mmep.090213
34. Hisham A. Kholidy, “An Intelligent Swarm based Prediction Approach for Predicting Cloud Computing User Resource Needs”, the Computer Communications Journal, December 2019.
35. Hisham A. Kholidy, “Towards A Scalable Symmetric Key Cryptographic Scheme: Performance Evaluation and Security Analysis”, IEEE Int. Conference on Computer Applications & Information Security (ICCAIS), Riyadh, Saudi Arabia, May 1-3, 2019.
36. Samar SH. Haytamy, Hisham A. Kholidy, Fatma A. Omara, “Integrated Cloud Services Dataset”, Lecture Note in Computer Science, ISBN 978-3-319-94471-5, https://doi.org/10.1007/978-3-319-94472-2. 14th World Congress on Services, 18-30. Held as Part of the Services Conf. Federation, SCF 2018, Seattle, WA, USA.
37. Hisham A. Kholidy, Ali T., Stefano I., Shamik S., Qian C., Sherif A., John H., “Attacks Detection in SCADA Systems Using an Improved Non- Nested Generalized Exemplars Algorithm", the 12th IEEE Int. Conf. on Computer Engineering and Systems (ICCES 2017), published in February 2018.

38. H. A. Kholidy and F. Baiardi, "CIDD: A Cloud Intrusion Detection Dataset for Cloud Computing and Masquerade Attacks," *2012 Ninth Int. Conference on Information Technology - New Generations*, 2012, pp. 397-402, doi: 10.1109/ITNG.2012.97.
39. Differentially Private Stochastic Gradient Descent. https://medium.com/pytorch/differential-privacy-series-part-1-dp-sgd-algorithm-explained-12512c3959a3